\documentclass[11pt,3p]{elsarticle}

\usepackage{graphicx}
\usepackage{amsmath}
\usepackage{amssymb}
\usepackage{url}
\usepackage{hyperref}
\usepackage[english]{babel}
\usepackage{color}
\usepackage{wrapfig}

\journal{Nuclear Physics A}

\begin{document}

\begin{frontmatter}

\title{Nuclear PDFs in the beginning of the LHC era}
\author{Hannu Paukkunen}
\ead{hannu.paukkunen@jyu.fi}
\address{Department of Physics, University of Jyv\"askyl\"a, P.O. Box 35, FI-40014, Finland}
\address{Helsinki Institute of Physics, University of Helsinki, P.O. Box 64, FI-00014, Finland}

\begin{abstract}
The status of the global fits of nuclear parton distributions (nPDFs) is reviewed. In addition to 
comparing the contemporary analyses of nPDFs, difficulties and controversies posed by
the neutrino-nucleus deeply inelastic scattering data is overviewed. At the end, the
first dijet data from the LHC proton+lead collisions is briefly discussed.

\end{abstract}

\end{frontmatter}

\section{Introduction}

\begin{figure}[tbhp]
\centering
\includegraphics[width=0.70\textwidth]{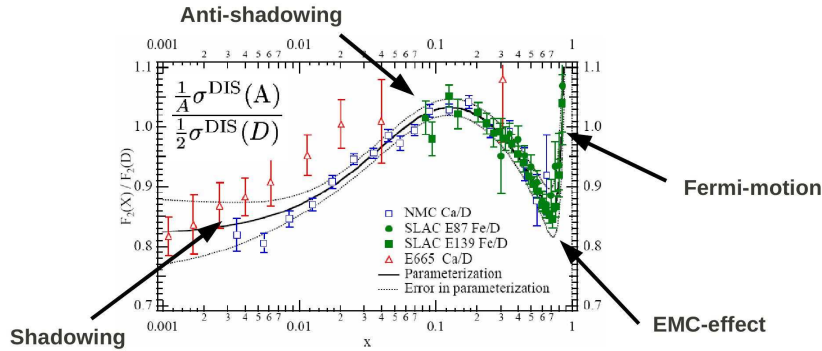}
\caption{Typical nuclear effects seen in the DIS measurements.} 
\label{fig:neffects}
\end{figure}

The experimental evidence for the appearance of non-trivial nuclear modifications in hard-process
cross sections is nowadays well known. The ``canonical'' example is the deeply inelastic scattering (DIS), in
which the ratio $\sigma(\ell^\pm+{\rm nucleus})/\sigma(\ell^\pm+{\rm deuteron})$ displays
the typical pattern of nuclear effects \cite{Arneodo:1992wf}: small-$x$ shadowing, antishadowing,
EMC-effect, and Fermi motion. A cartoonic picture is shown in Fig.~\ref{fig:neffects}.
The central theme in the global analyses of nuclear parton distributions $f_i^A$ (nPDFs), is to find out whether, and to what extent
(in which processes, in which kinematic conditions) such effects can be interpreted in terms of standard
collinear factorization \cite{Collins:1989gx,Collins:1985ue}, for example, in the case of DIS,
\begin{equation}
 \sigma_{\rm DIS}^{\ell+A} = \sum_i 
  \mathop{\mathop{\underbrace{f_i^A(\mu_{\rm fact}^2)}}_
 {\rm nuclear \,\, PDFs, \,\, obey}}_
 {\rm the \,\, usual \,\, DGLAP}
 \otimes \,\,\,\,\,\, 
  \mathop{\mathop{\underbrace{\hat \sigma_{\rm DIS}^{\ell+i}(\mu_{\rm fact}^2,\mu_{\rm ren}^2)}}_
 {\rm usual \,\, pQCD}}_
 {\rm coefficient \\ \,\, functions} \,\, + \,\, \mathcal{O} \left({1}/{Q^n}\right),
 \label{eq:1}
\end{equation}
where the partonic coefficient functions $\hat \sigma$ and the DGLAP evolution \cite{Dokshitzer:1977sg,Gribov:1972ri,Gribov:1972rt,Altarelli:1977zs}
of $f_i^A$ are the same as in the case of free proton scattering.
In short, the goal is to carry out very similar program as
in the case of free proton analyses \cite{Forte:2013wc,Martin:2009iq,Lai:2010vv,Ball:2012cx}.
The main obstacle in drawing definite conclusions regarding the adequacy of factorization in nuclear environment has
been the shortage of suitable hard-process data as, in comparison
to the free proton case, the amount, kinematic reach, and variety of experimental data is rather restricted.
In fact, it is only very recently --- especially along the discussion surrounding the neutrino-nucleus DIS (see later)
--- that the nPDF process independence has been put to a serious test. In the near future, the LHC proton-lead
hard-process data \cite{Loizides:2013nka}, once available in their full glory, are expected to provide definitive answers.
The theoretical expectations are that the $Q^{-n}$-type power corrections in Eq.~(\ref{eq:1})
become enhanced in nuclear environment and could eventually be seen even at perturbative scales 
\cite{Eskola:2003gc}. The search for such a transition from the linear DGLAP dynamics to a non-linear
regime is also one of the main goals of the LHC proton+lead program \cite{Albacete:2014fwa,Salgado:2011wc}.

\vspace{-0.3cm}
\section{A brief overview of the existing analyses of nuclear PDFs}

The latest available next-to-leading order (NLO) nPDFs are \textsc{hkn07} \cite{Hirai:2007sx},
\textsc{eps09} \cite{Eskola:2009uj}, and \textsc{dssz} \cite{deFlorian:2011fp}. Also the work of nCTEQ 
collaboration \cite{Schienbein:2009kk,Kovarik:2010uv,Kovarik:2013sya}
will be commented, although their final pre-LHC parametrization is not yet available.
Some characteristics of these NLO fits are summarized in Table~\ref{tab:nPDFsummary}.

\begin{table*}[tbh]
\caption{Key characteristics of the contemporary nPDF fits (the nCTEQ column corresponds to the presentation
given in DIS2013, Marseille \cite{Kovarik:2013sya}).
}
\begin{center}
\begin{tabular}{|c||c|c|c|c|}
\hline
& \textsc{hkn07}  & \textsc{eps09}	 & \textsc{dssz}	 	& \textsc{ncteq} \\
\hline
\small{Order in $\alpha_s$}                   & \small{LO} \& \small{NLO} & \small{LO \& NLO}           & \small{NLO}               & \small{NLO}         \\
\small{Neutral current DIS $\ell$+A/$\ell$+d} & \checkmark                & \checkmark                  & \checkmark                & \checkmark          \\
\small{Drell-Yan dilepton p+A/p+d}            & \checkmark                & \checkmark                  & \checkmark                & \checkmark          \\
\small{RHIC pions d+Au/p+p}                   &                           & \checkmark                  & \checkmark                &                     \\
\small{Neutrino-nucleus DIS}                  &                           &                             & \checkmark                &                     \\
\small{$Q^2$ cut in DIS}                      & $1 \, {\rm GeV}$          & $1.3 \, {\rm GeV}$          & $1 \, {\rm GeV}$          & $2 \, {\rm GeV}$    \\
\small{datapoints}                            & 1241                      & 929                         & 1579                      & 708                 \\
\small{free parameters}                       & 12                        & 15                          & 25                        & 17                  \\
\small{error analysis}                        & \checkmark                & \checkmark                  & \checkmark                & \checkmark          \\
\small{error tolerance $\Delta \chi^2$}       & 13.7                      & 50                          & 30                        & 35                  \\
\small{Free proton baseline PDFs}             & \small{\textsc{mrst98}}   & \small{\textsc{cteq6.1}}    & \small{\textsc{mstw2008}} & \small{\textsc{cteq6m}-like} \\
\small{Heavy quark treatment}                 & ZM-VFNS                   & ZM-VFNS                     & GM-VFNS                   & GM-VFNS             \\
\hline
\end{tabular}
\end{center}
\label{tab:nPDFsummary}
\end{table*}

The nPDFs $f_i^A$ are linear combinations of bound proton ($f_i^{\rm p,A}$) and bound neutron
($f_i^{\rm n,A}$) PDFs
\begin{equation}
 f_i^{A}(x,Q^2) = \left(\frac{Z}{A}\right) f_i^{\rm p,A}(x,Q^2) + \left(\frac{N}{A}\right) f_i^{\rm n,A}(x,Q^2),
\end{equation}
where $Z$ is the number of protons and $N$ the number of neutrons in the nucleus $A$. The relation
between the bound proton PDFs and the free nucleon baseline $f_i^p$ is usually expressed as
\begin{equation}
 f_i^{p,A}(x,Q^2) = R^A_i(x,Q^2) f_i^p(x,Q^2),
\end{equation}
where $R^A_i(x,Q^2)$ quantifies the nuclear modification (also impact-parameter dependent versions
has been suggested, see Ref.~\cite{Helenius:2012wd}). For the moment, all groups rely
on the isospin symmetry to obtain the bound neutron PDFs (e.g. $f_u^{n,A}=f_d^{p,A}$) --- an assumption
that would need to be revised once the QED effects are included in the parton evolution
\cite{Martin:2004dh,Ball:2013hta,Bertone:2013vaa,Sadykov:2014aua}.
All but \textsc{hkn07} assume no nuclear modification for the deuteron, $R^{\rm deuteron}_i(x,Q^2) = 1$.
Although small, the nuclear effects in deuteron are still non-zero, and have some importance when the
deuteron data are included in the free proton fits \cite{Martin:2012da}.

Different groups use different functions to parametrize $R^A_i(x,Q_0^2)$. For example, while \textsc{eps09} employs
a piecewize fit function (as a function of $x$), \textsc{dssz} uses a single fit function constructed such that the 
analytic Mellin transform exists. In the works of nCTEQ, $f_i^{p,A}(x,Q_0^2)$ is parametrized
directly with the same fit function as used for their free proton baseline. However, as the
free proton baseline is taken as ``frozen'', this is simply another way of parametrizing $R^A_i(x,Q_0^2)$.

Most of the data that are used as constraints in the nPDF fits come as nuclear ratios similar to that shown
in Fig.~\ref{fig:neffects}. What makes such ratios especially appealing is that they prove remarkably inert
to the higher order pQCD corrections. Also, the dependence of the free proton baseline PDFs gets reduced.
The exception here are the neutrino-nucleus DIS data, included in the \textsc{dssz} fit, that are only available as absolute
cross-sections (or as corresponding structure functions derived from those). The inclusion of these data 
also requires using a general-mass variable-flavor-number scheme (GM-VFNS) for treating the heavy quarks overtaking the zero-mass
scheme (ZM-VFNS) employed in the older fits (\textsc{eps09}, \textsc{hkn07}).

\begin{figure}[tbhp]
\centering
\includegraphics[width=0.49\textwidth]{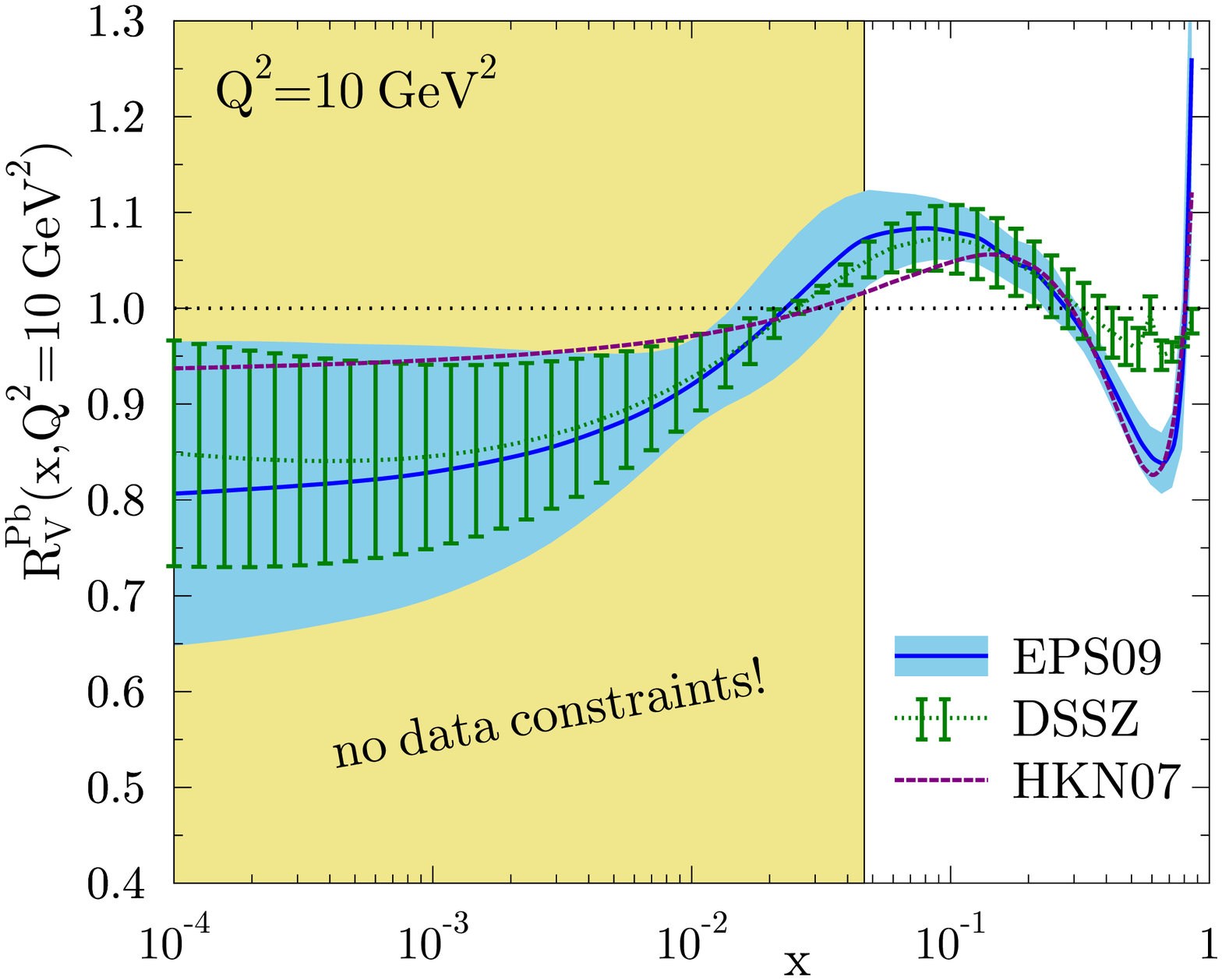}
\includegraphics[width=0.49\textwidth]{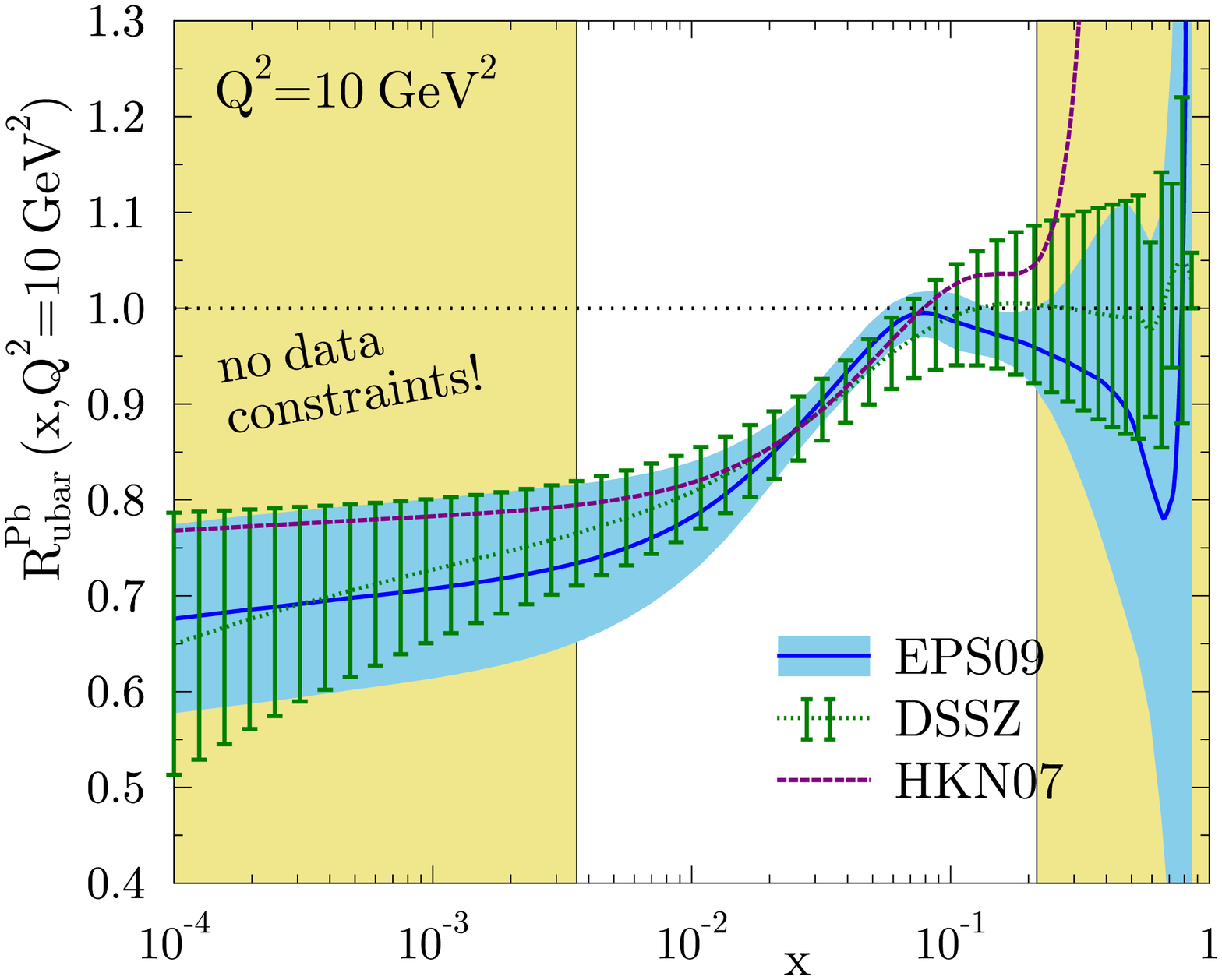}
\caption{Comparison of up valence and sea quark nuclear modification factors for the lead nucleus
at $Q^2 = 10 \, {\rm GeV}^2$. Blue line with error band is \textsc{eps09}, green dotted line with error bars  \textsc{dssz}, 
and purple dashed  \textsc{hkn07}.} 
\label{fig:Valence_Sea}
\end{figure}

A comparison of the $R^{\rm Pb}_{u_{V}}(x,Q^2=10 \, {\rm GeV}^2)$ (up valence) and $R^{\rm Pb}_{\overline u}(x,Q^2=10 \, {\rm GeV}^2)$ (up sea)
from the available parametrizations is presented in Fig.~\ref{fig:Valence_Sea}. The areas with yellow background
are those regions of $x$ where the direct data constraints do not exist or they 
are very weak. In these regions the bias due to the assumed form of the fit function and parameter fixing
may be significant. Whereas the $R^A_{u_{V}}$ from
\textsc{eps09} and \textsc{hkn07} agree at large $x$, \textsc{dssz}, strangely enough, is clearly above at $x\simeq 0.5$. This is
rather unexpected as in this EMC region there are plenty of data constraints from DIS experiments. 
The same behaviour is there already in the \textsc{dssz} precursor, \textsc{nds} \cite{deFlorian:2003qf}, and the
probable source of this has been identified as a misinterpretation of the isospin correction
that the experiments have applied to the data\footnote{M. Stratmann and P. Zurita, priv.comm.}.
In \textsc{eps09} and \textsc{hkn07} the assumption $R^A_{u_V}(x,Q^2_0) = R^A_{d_V}(x,Q^2_0)$ was made
as only one type of data sensitive to the large-$x$ valence quarks was included in these fits.
Indeed, at large $x$, one can approximate
\begin{equation}
 d\sigma_{\rm DIS}^{\ell+A} \propto \left(\frac{4}{9}\right)u_V^A + \left(\frac{1}{9}\right) d_V^A
 \propto u_V^p \left[ R^A_{u_V} + R^A_{d_V} \frac{d_V^p}{u_V^p} \frac{Z+4N}{N+4Z} \right]
 \approx u_V^p \left[R^A_{u_V} + \frac{1}{2} R^A_{d_V} \right],
\end{equation}
which underscores the fact that these data can constrain only a certain linear combination
of $R^A_{u_V}$ and $R^A_{d_V}$. Despite the lack of other type of data sensitive to the valence quarks, the assumption
$R^A_{u_V}(x,Q^2_0) = R^A_{d_V}(x,Q^2_0)$ was released in a recent nCTEQ work leading to
mutually wildly different $R^A_{u_V}$ and $R^A_{d_V}$ (see Fig.1 in Ref.\cite{Kovarik:2013sya}).
Other type of data sensitive to the valence quarks would obviously be required to pin
down them separately in a more realistic manner. Despite the fact that some neutrino data
(also sensitive to the valence quarks) was included in the \textsc{dssz} fit, the authors
did not investigate the possible difference between $R^A_{u_V}$ and $R^A_{d_V}$.

In the case of $R^A_{\overline u}$, which here generally represents the sea quark modification,
all parametrizations are in a fair agreement in the data-constrained region.
This is also true if the nCTEQ results are considered (Fig.1 in Ref.\cite{Kovarik:2013sya}).
Above the parametrization scale $Q^2 > Q^2_0$, the sea quark modifications are also significantly affected,
especially at large $x$ ($x\gtrsim 0.2$), by the corresponding gluon modification $R^A_{g}$ via
the DGLAP evolution.

\begin{figure}[tbhp]
\centering
\includegraphics[width=0.49\textwidth]{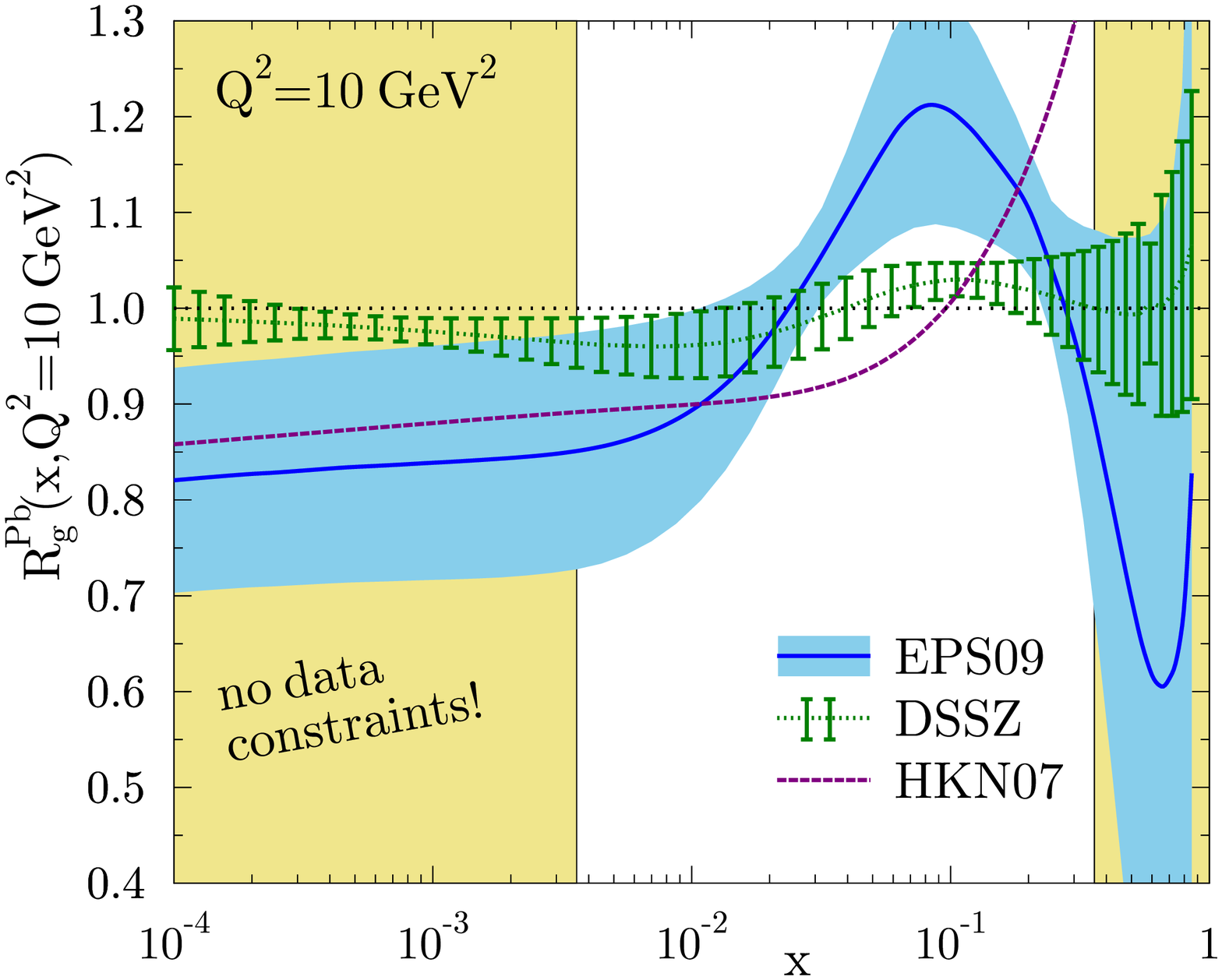}
\includegraphics[width=0.49\textwidth]{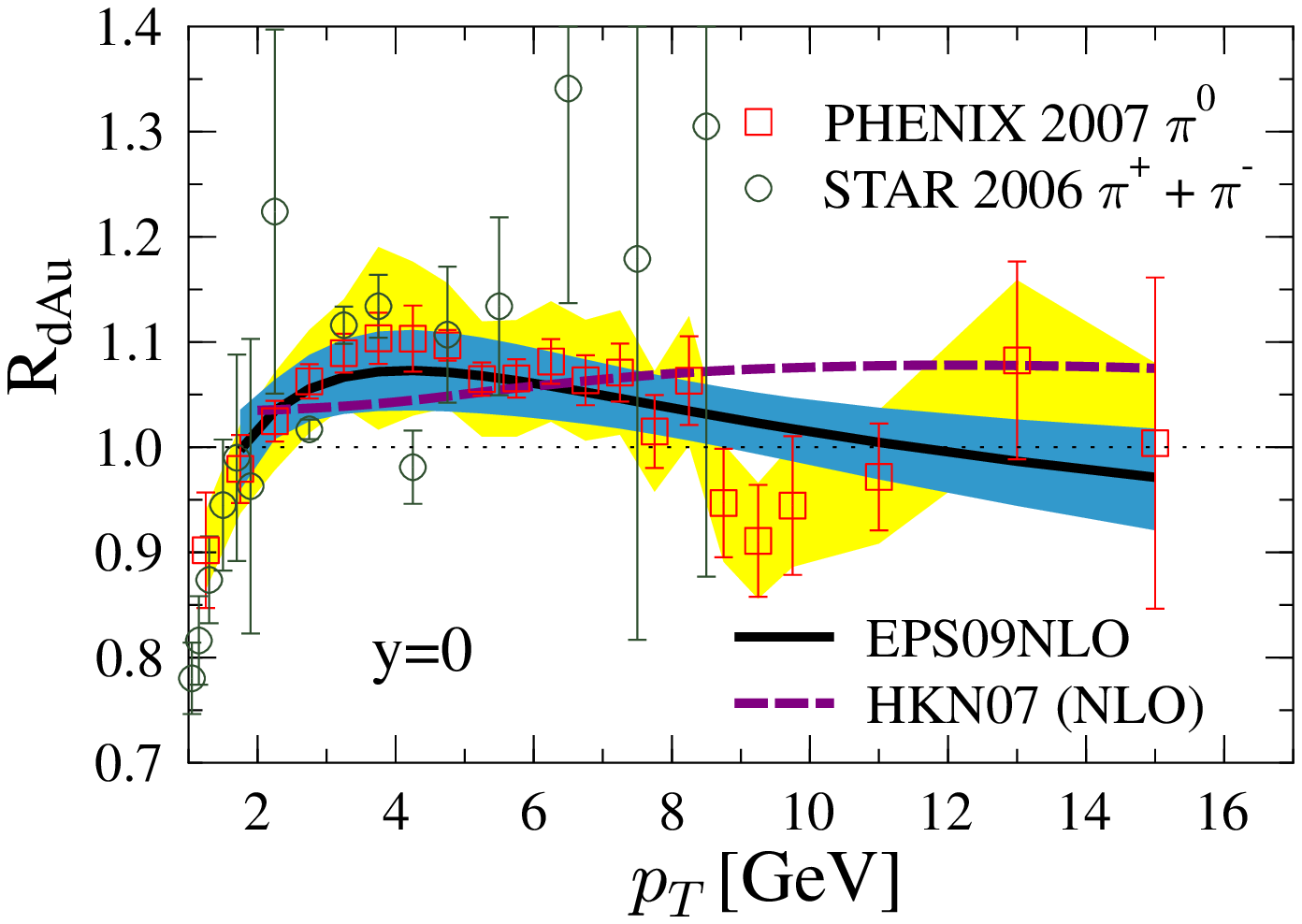}
\caption{Comparison of the gluon nuclear modification factors for the lead nucleus at $Q^2 = 10 \, {\rm GeV}^2$ (left),
and the nuclear modification for inclusive pion production in d+Au collisions at midrapidity (right).} 
\label{fig:G}
\end{figure}

The largest differences among \textsc{eps09}, \textsc{hkn07}, and \textsc{dssz} are in the nuclear
effects for the gluon PDFs, shown in Fig.~\ref{fig:G}. The origins of the large differences
are more or less known: The DIS and Drell-Yan data are mainly sensitive to the quarks, and thus leave 
$R^A_g$ quite unconstrained.
To improve on this, \textsc{eps09} and \textsc{dssz} make use of the nuclear modification observed in the 
inclusive pion production at RHIC \cite{Abelev:2009hx,Adler:2006wg}. An example of
these data are shown in Fig.~\ref{fig:G}. Although the pion data included in \textsc{eps09} and \textsc{dssz} are not
exactly the same, it may still look surprising how different the resulting $R^A_g$ are. The reason
lies (as noted also e.g. in \cite{Eskola:2012rg}) in the use of different parton-to-pion
fragmentation functions (FFs) $D^{k\rightarrow \pi + X}(z,Q^2)$ in the calculation of the inclusive 
pion production cross sections
\begin{equation}
 d\sigma^{\rm {d+Au} \rightarrow \pi + X} = \sum_{i,j,k}f_i^{\rm d} \otimes d\hat \sigma^{ij\rightarrow k} \otimes f_j^{\rm Au} \otimes D^{k\rightarrow \pi + X}.
\end{equation}
While in \textsc{eps09} the \textsc{kkp} \cite{Kniehl:2000fe} vacuum FFs (determined solely from the $e^+e^-$ data)
were used, \textsc{dssz} advocated the use of nuclear modified FFs \cite{Sassot:2009sh} determined
from the very \emph{same} pion data that was later on included in the \textsc{dssz} nPDF fit. Therefore, they
were condemned to find a very similar $R^A_g$ that was used in the fit for the nuclear modified FFs,
namely that of \textsc{nds}. While both \textsc{eps09} and \textsc{dssz} can describe the 
pion data, the physics content is rather different (initial state effect in \textsc{eps09}, final state effect in \textsc{dssz}).
The nCTEQ result for the gluons (Fig.1 in Ref.\cite{Kovarik:2013sya}) comes with clearly larger uncertainty than the ones in
\textsc{eps09} or \textsc{dssz}. This is mainly due to the larger $Q^2$ cut for the DIS data and that they do
not currently include the pion data.

\vspace{-0.3cm}
\section{The case of neutrino-nucleus DIS data}

A nPDF-related issue that has caused some stir during the recent years concerns the the compatibility of
the neutrino-nucleus DIS data with the other nuclear data (used e.g. in \textsc{eps09}).
The whole discussion was initiated in \cite{Schienbein:2007fs} where a PDF fit to the NuTeV neutrino+iron 
DIS data \cite{Tzanov:2005kr} was reported. The results seemed to point towards nuclear modifications in PDFs different from those
obtained in charged-lepton-induced reactions. Later on, it was argued \cite{Kovarik:2010uv} that
the these two types of data display clear mutual tension, and, even a breakdown of the collinear
factorization was flashed as a possible explanation. Such would have far reaching consequences as 
the all major free proton groups \cite{Martin:2009iq,Lai:2010vv,Ball:2012cx} do include neutrino data in their fits, thereby
silently assuming the validity of the factorization there. The authors underscored accounting for 
the NuTeV data correlations via the provided covariance matrix although the same conclusion was reached
when all the errors were added in quadrature. 

\begin{figure}[tbhp]
\centering
\includegraphics[width=0.70\textwidth]{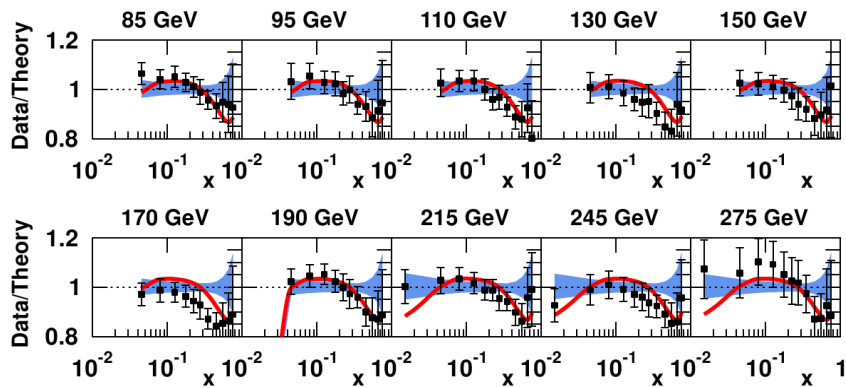}
\caption{An example of the $Q^2$-averaged nuclear modifications derived from the NuTeV data. The data points
correspond to the data divided by NLO calculations with \textsc{cteq6.6} PDFs without nuclear effects, and the blue band
is the \textsc{cteq6.6} uncertainty band. The red lines are predictions based on \textsc{eps09}. Each panel corresponds
to a different neutrino beam energy.} 
\label{fig:nutev}
\end{figure}

In \cite{Paukkunen:2010hb} a somewhat different strategy was adopted. Instead of concentrating on
one data set only, this study aimed for a more global picture by using also neutrino
data from the CHORUS \cite{Onengut:2005kv} and CDHSW \cite{Berge:1989hr} experiments. By confronting
the existing sets of nuclear PDFs with these data, no obvious difficulty to reproduce them was
detected --- the overall agreement seemed to be good. However, the NuTeV cross sections were
found to contain rather large, neutrino-energy dependent fluctuations.
Such fluctuations were found upon inspecting the nuclear modifications 
$\sigma(\rm experimental)/\sigma(\rm theory)$ as a function of $x$, averaged over $Q^2$.
Examples of such ratios in the case of NuTeV data are shown in Fig.~\ref{fig:nutev}.
Unexpected fluctuations between different neutrino beam energies are clearly visible.

A solution to all this was proposed in \cite{Paukkunen:2013grz}. The central idea was
to look (for each neutrino beam energy $E_\nu$ separately) the cross sections normalized by 
the total cross section (integrated over $x$ and $Q^2$) --- nothing more
extraordinary than e.g. measuring the shape of the dilepton distributions at the LHC \cite{Chatrchyan:2011wt}
or Tevatron \cite{Abazov:2007jy}.
\begin{figure*}[ht]
\begin{minipage}[b]{0.45\linewidth}
\centering
\includegraphics[width=\textwidth]{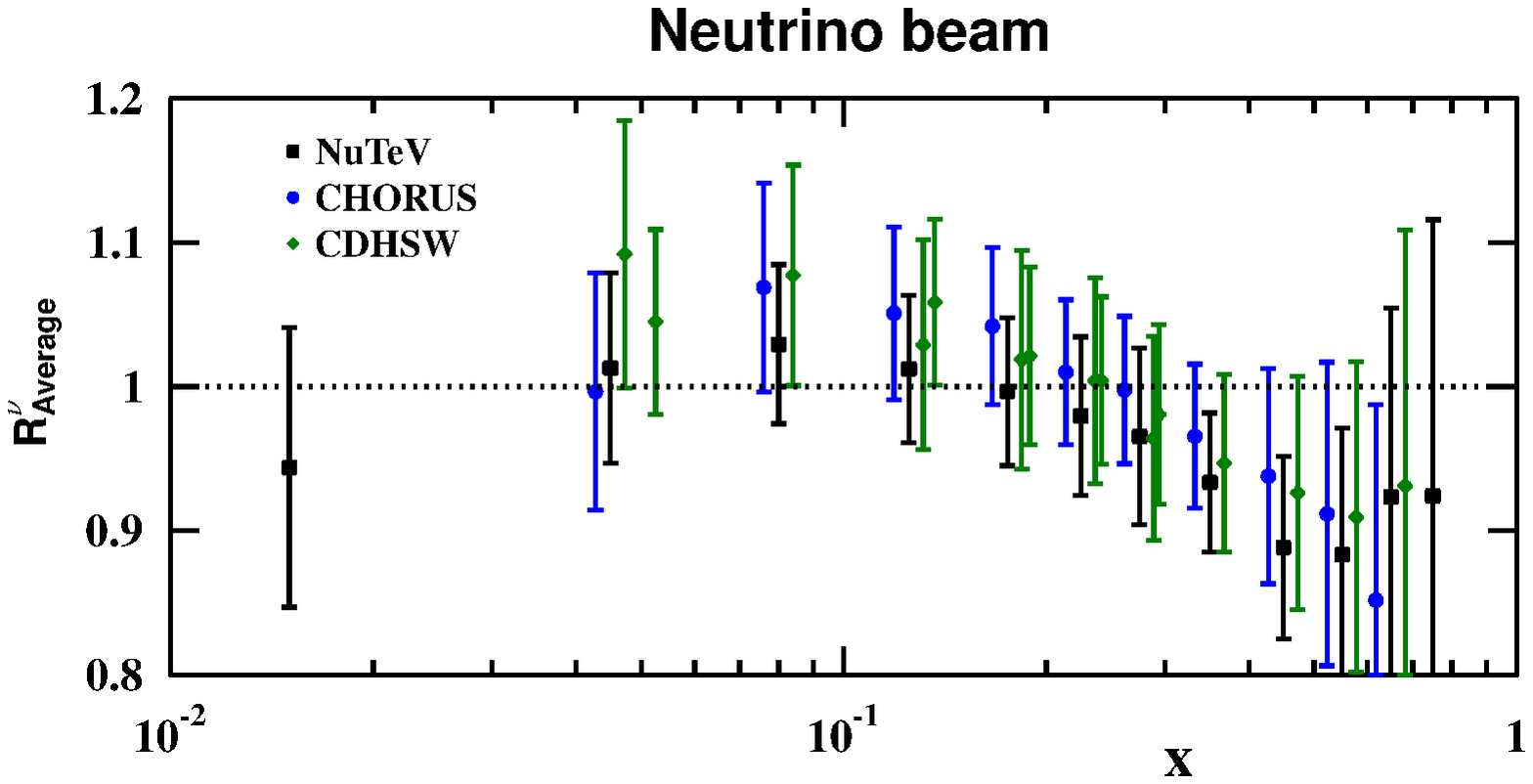}
\end{minipage}
\hspace{0.5cm}
\begin{minipage}[b]{0.45\linewidth}
\centering
\includegraphics[width=\textwidth]{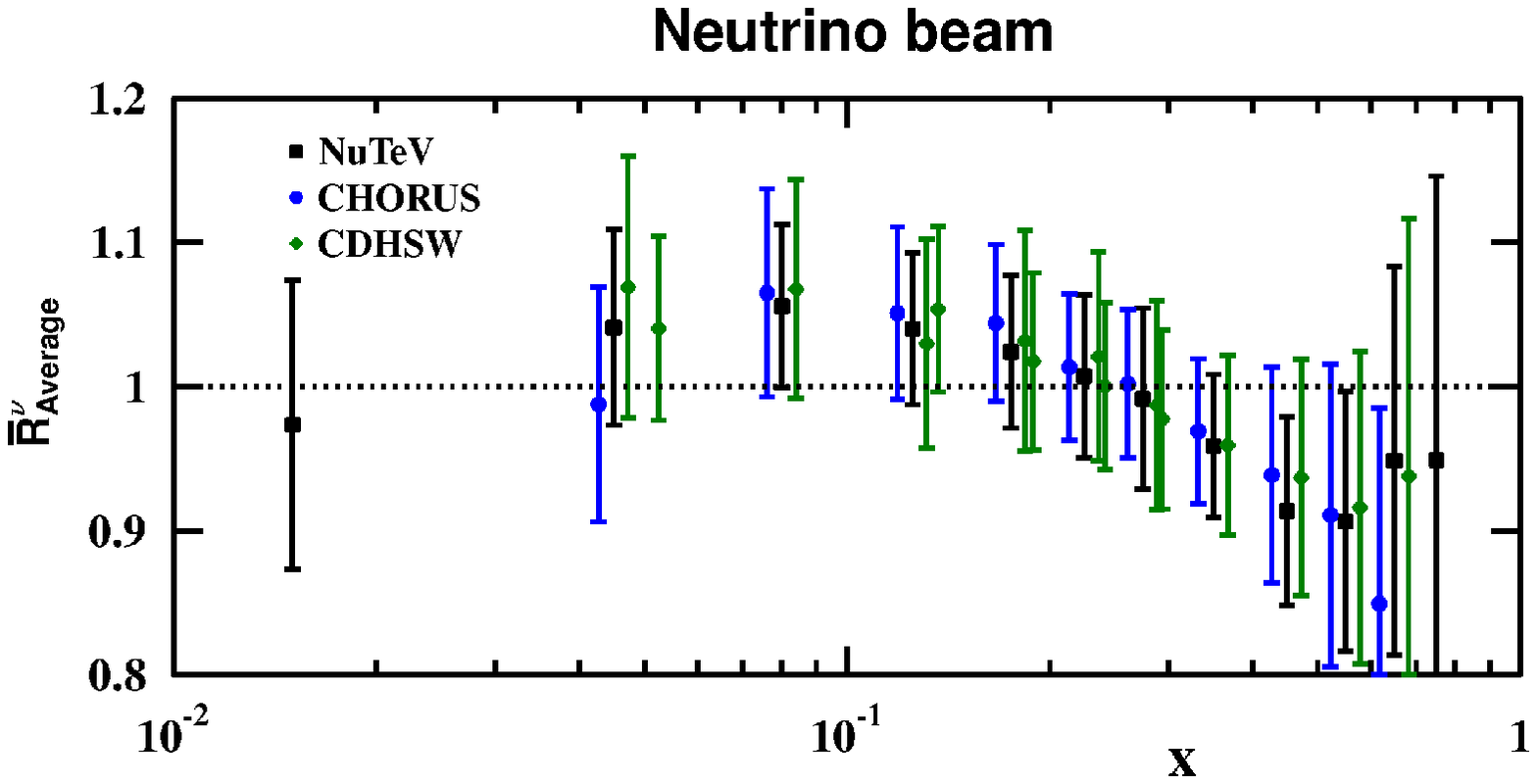}
\end{minipage}
\caption{Neutrino data (averaged over $Q^2$ and $E_\nu$) without the normalization (left-hand panel),
and with the normalization (right-hand panel). The CHORUS (blue circles) and CDHSW (green diamonds)
data has been horizontally slightly shifted from the NuTeV (black squares) data points.}
\label{Fig:Neutrino1}
\end{figure*}
Fig.~\ref{Fig:Neutrino1} illustrates the effect, presenting the 
nuclear modifications (averaged over $Q^2$ and $E_\nu$), first, obtained without the normalization (left), and then, after applying
the normalization procedure (right). The normalization process definitely improves the
mutual agreement among the independent data sets, and implies that, apart from the normalization,
the $x$ and $Q^2$ dependence of all the data appear to agree.
The compatibility of these data with the nuclear PDFs (\textsc{cteq6.6}+\textsc{eps09}) was studied employing a novel re-weighting
technique (reminiscent of similar methods developed in Refs. \cite{Ball:2010gb,Ball:2011gg,Watt:2012tq}). 
While supporting the compatibility, the method also confirmed the same incompatibility
as found in the nCTEQ work when the normalization was not applied.

A third point-of-view is provided by the authors of \textsc{dssz} who actually went and included neutrino data
(also NuTeV data) in their nPDF fit without an obvious difficulty. This may seem surprising now that the difficulties
posed by the NuTeV data appear confirmed. Three issues contributing can be readily identified:
the use of structure functions instead of the absolute cross sections (much less data), the use
of \textsc{mstw2008} PDFs \cite{Martin:2009iq}
as a baseline (already constrained by the NuTeV data), and adding the \textsc{mstw2008} PDF uncertainties on top of the
experimental errors. All this makes the neutrino data less important in comparison to the other
data and may thereby obscure the difficulties that the NuTeV data poses.

\vspace{-0.3cm}
\section{The first glimpse of nuclear PDFs at the LHC?}

\begin{figure}[ht!]
\centerline{\includegraphics[width=0.5\textwidth]{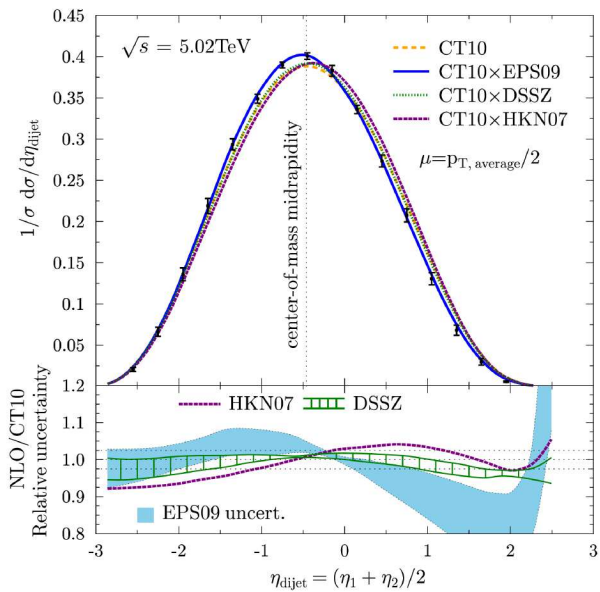}}
\caption[]{The preliminary dijet rapidity distribution from CMS (read ``by eye`` from \cite{Chatrchyan:2014hqa}) 
compared to the calculations \cite{Eskola:2013aya} with no nuclear effects (orange dotted),
with \textsc{eps09} (solid blue), \textsc{hkn07} (purple long dashed), and \textsc{dssz} (green dotted).}
\label{fig:dijet}
\end{figure}

The first experimental results from the LHC proton+lead run are now starting to become public \cite{Loizides:2013nka},
and are expected to conclusively test the universality of the nPDFs and hopefully serve as further constraints as well.
For instance, different rapidity and transverse momentum distributions of charged leptons from electroweak gauge boson
decays have been theoretically explored \cite{Armesto:2013kqa,Kang:2012am,Brandt:2014vva,Guzey:2012jp,Paukkunen:2010qg}
and promise to probe various aspects of nPDFs. Direct photons, inclusive hadrons and jet observables
\cite{Arleo:2007js,Stavreva:2010mw,Arleo:2011gc,Albacete:2013ei,He:2011sg,Dai:2013xca} have been predicted to retain sensitivity 
to the nuclear modifications in PDFs as well. However,
the first hard QCD-process data set from the 2013 proton+lead run that can be directly compared to the nPDF
predictions at the NLO level is the shape of the pseudorapidity distribution of dijets measured by 
the CMS collaboration \cite{Chatrchyan:2014hqa}. The dijet pseudorapidity $\eta_{\rm dijet}$ is
defined as the average of the leading and subleading jet pseudorapidities,
\begin{equation}
\eta_{\rm dijet} \equiv (\eta_{\rm leading} + \eta_{\rm subleading})/2,
\end{equation}
within the acceptance $|\eta_{\rm leading,subleading}| < 3$. The transverse momenta of the jets
are large, $p_{T, {\rm leading}} > 120 \, {\rm GeV}$, $p_{T, {\rm subleading}} > 30 \, {\rm GeV}$,
and the rather narrow cone size $R=0.3$ used in defining the jet within the anti-$k_T$ algorithm,
suppresses the uncertainties from the underlying event. 
The corresponding pQCD predictions appeared in Ref.~\cite{Eskola:2013aya} where it was observed that this
particular observable --- not being an absolute cross section but a normalized one --- is surprisingly
inert to the higher order pQCD corrections but still retains sensitivity to the nuclear modifications in PDFs. 
These predictions are, in Fig.~\ref{fig:dijet}, compared to the preliminary data taken from Ref.~\cite{Chatrchyan:2014hqa}.
In comparison to the calculation with just \textsc{ct10} PDFs with no nuclear effects (orange
dashed line) the preliminary data shows a clear enhancement in the backward direction
and suppression in the forward direction. These are readily explained by an antishadowing
and EMC effect in the gluon PDFs, similar to those in \textsc{eps09}, as the good agreement with
the preliminary data and the calculation with \textsc{eps09} (blue solid curve)
demonstrates. The nuclear effects in \textsc{dssz} are too weak to reproduce the data (green dotted line),
and \textsc{hkn07} makes a correction to a wrong direction (purple dashed line).
The bottom line here is that the same nuclear modifications required to reproduce the inclusive pion
data (Fig.~\ref{fig:G}) at $\sqrt{s}=200 \, {\rm GeV}$ at rather low transverse momenta, $p_T < 20 \, {\rm GeV}$,
are supported by these preliminary dijet data with $\sqrt{s}=5.02 \, {\rm TeV}$ and $p_T \gtrsim 100 \, {\rm GeV}$.
This is evidence of the nPDF universality and also seems to indicate that there cannot be significant
nuclear modifications in the parton-to-hadron FFs like those proposed in \cite{Sassot:2009sh}.

The role of nPDFs in the LHC lead+lead program is not as evident as it is in proton+lead collisions.
However, it is encouraging that even in this
case the existing measurements for high-$p_T$ electroweak observables are consistent with the pQCD predictions
\cite{Chatrchyan:2012vq,Chatrchyan:2012nt,Chatrchyan:2011ua,Aad:2012ew}, the data uncertainties being admittedly
still rather large. Such observations give further confidence to the use of collinear factorization e.g. in
computing the initial condition needed for subsequent hydro-dynamical evolution \cite{Paatelainen:2013eea},
or in estimating the backgrounds when charting different sources of low-$p_T$ photons \cite{Klasen:2013mga,Chatterjee:2013naa} 
in heavy-ion collisions.

\vspace{-0.4cm}
\section{Summary}

The present status of the nPDFs was reviewed. Sizable differences among the independent
parametrizations exist reflecting the significant prevailing uncertainty. To definitely
reduce these uncertainties, more data are needed. In this respect, data from the LHC proton-lead
run are foreseen to bring significant additional insight. Indeed, already the very first
dijet data from the CMS collaboration, briefly mentioned here, could help to discriminate
between the existing sets of nPDFs. The long-standing issue concerning the neutrino-nucleus
DIS was also shortly recalled for which a possible solution has been recently proposed and
could eventually lead to a more comprehensive use of these data in global fits of nPDFs. 

\vspace{-0.4cm}
\section*{Acknowledgments}
\noindent Financial support from the Academy of Finland, Project No. 133005, is acknowledged.


\vspace{-0.4cm}
\begin{thebibliography}{99}

%\cite{Arneodo:1992wf}
\bibitem{Arneodo:1992wf}
  M.~Arneodo,
  %``Nuclear effects in structure functions,''
  Phys.\ Rept.\  {\bf 240} (1994) 301.
  %%CITATION = PRPLC,240,301;%%
  %366 citations counted in INSPIRE as of 06 Jan 2014

%\cite{Collins:1989gx}
\bibitem{Collins:1989gx}
  J.~C.~Collins, D.~E.~Soper and G.~F.~Sterman,
  %``Factorization of Hard Processes in QCD,''
  Adv.\ Ser.\ Direct.\ High Energy Phys.\  {\bf 5} (1988) 1
  [hep-ph/0409313].
  %%CITATION = HEP-PH/0409313;%%
  %412 citations counted in INSPIRE as of 06 Jan 2014

%\cite{Collins:1985ue}
\bibitem{Collins:1985ue}
  J.~C.~Collins, D.~E.~Soper and G.~F.~Sterman,
  %``Factorization for Short Distance Hadron - Hadron Scattering,''
  Nucl.\ Phys.\ B {\bf 261} (1985) 104.
  %%CITATION = NUPHA,B261,104;%%
  %360 citations counted in INSPIRE as of 06 Jan 2014

%\cite{Dokshitzer:1977sg}
\bibitem{Dokshitzer:1977sg}
  Y.~L.~Dokshitzer,
  %``Calculation of the Structure Functions for Deep Inelastic Scattering and e+ e- Annihilation by Perturbation Theory in Quantum Chromodynamics.,''
  Sov.\ Phys.\ JETP {\bf 46} (1977) 641
   [Zh.\ Eksp.\ Teor.\ Fiz.\  {\bf 73} (1977) 1216].
  %%CITATION = SPHJA,46,641;%%
  %2361 citations counted in INSPIRE as of 06 Jan 2014

%\cite{Gribov:1972ri}
\bibitem{Gribov:1972ri}
  V.~N.~Gribov and L.~N.~Lipatov,
  %``Deep inelastic e p scattering in perturbation theory,''
  Sov.\ J.\ Nucl.\ Phys.\  {\bf 15} (1972) 438
   [Yad.\ Fiz.\  {\bf 15} (1972) 781].
  %%CITATION = SJNCA,15,438;%%
  %2709 citations counted in INSPIRE as of 06 Jan 2014

%\cite{Gribov:1972rt}
\bibitem{Gribov:1972rt}
  V.~N.~Gribov and L.~N.~Lipatov,
  %``e+ e- pair annihilation and deep inelastic e p scattering in perturbation theory,''
  Sov.\ J.\ Nucl.\ Phys.\  {\bf 15} (1972) 675
   [Yad.\ Fiz.\  {\bf 15} (1972) 1218].
  %%CITATION = SJNCA,15,675;%%
  %1026 citations counted in INSPIRE as of 06 Jan 2014

%\cite{Altarelli:1977zs}
\bibitem{Altarelli:1977zs}
  G.~Altarelli and G.~Parisi,
  %``Asymptotic Freedom in Parton Language,''
  Nucl.\ Phys.\ B {\bf 126} (1977) 298.
  %%CITATION = NUPHA,B126,298;%%
  %4841 citations counted in INSPIRE as of 06 Jan 2014

%\cite{Forte:2013wc}
\bibitem{Forte:2013wc}
  S.~Forte and G.~Watt,
  %``Progress in the Determination of the Partonic Structure of the Proton,''
  Ann.\ Rev.\ Nucl.\ Part.\ Sci.\  {\bf 63} (2013) 291
  [arXiv:1301.6754 [hep-ph]].
  %%CITATION = ARXIV:1301.6754;%%
  %23 citations counted in INSPIRE as of 22 Nov 2013
 
%\cite{Martin:2009iq}
\bibitem{Martin:2009iq}
  A.~D.~Martin, W.~J.~Stirling, R.~S.~Thorne and G.~Watt,
  %``Parton distributions for the LHC,''
  Eur.\ Phys.\ J.\ C {\bf 63} (2009) 189
  [arXiv:0901.0002 [hep-ph]].
  %%CITATION = ARXIV:0901.0002;%%
  %1859 citations counted in INSPIRE as of 06 Jan 2014

%\cite{Lai:2010vv}
\bibitem{Lai:2010vv}
  H.~-L.~Lai, M.~Guzzi, J.~Huston, Z.~Li, P.~M.~Nadolsky, J.~Pumplin and C.~-P.~Yuan,
  %``New parton distributions for collider physics,''
  Phys.\ Rev.\ D {\bf 82} (2010) 074024
  [arXiv:1007.2241 [hep-ph]].
  %%CITATION = ARXIV:1007.2241;%%
  %745 citations counted in INSPIRE as of 06 Jan 2014

%\cite{Ball:2012cx}
\bibitem{Ball:2012cx}
  R.~D.~Ball, V.~Bertone, S.~Carrazza, C.~S.~Deans, L.~Del Debbio, S.~Forte, A.~Guffanti and N.~P.~Hartland {\it et al.},
  %``Parton distributions with LHC data,''
  Nucl.\ Phys.\ B {\bf 867} (2013) 244
  [arXiv:1207.1303 [hep-ph]].
  %%CITATION = ARXIV:1207.1303;%%
  %95 citations counted in INSPIRE as of 06 Jan 2014

    %\cite{Loizides:2013nka}
\bibitem{Loizides:2013nka}
  C.~Loizides,
  %``First results from p-Pb collisions at the LHC,''
  EPJ Web Conf.\  {\bf 60} (2013) 06004
  [arXiv:1308.1377 [nucl-ex]].
  %%CITATION = ARXIV:1308.1377;%%
  %3 citations counted in INSPIRE as of 06 Jan 2014

%\cite{Eskola:2003gc}
\bibitem{Eskola:2003gc}
  K.~J.~Eskola, H.~Honkanen, V.~J.~Kolhinen, J.~-w.~Qiu and C.~A.~Salgado,
  %``Nonlinear corrections to the DGLAP equations: Looking for the saturation limits,''
  hep-ph/0302185.
  %%CITATION = HEP-PH/0302185;%%
  %12 citations counted in INSPIRE as of 02 Apr 2014
  
%\cite{Albacete:2014fwa}
\bibitem{Albacete:2014fwa}
  J.~L.~Albacete and C.~Marquet,
  %``Gluon saturation and initial conditions for relativistic heavy ion collisions,''
  Prog.\ Part.\ Nucl.\ Phys.\  {\bf 76} (2014) 1
  [arXiv:1401.4866 [hep-ph]].
  %%CITATION = ARXIV:1401.4866;%%
  %3 citations counted in INSPIRE as of 02 Apr 2014  

  %\cite{Salgado:2011wc}
\bibitem{Salgado:2011wc}
  C.~A.~Salgado, J.~Alvarez-Muniz, F.~Arleo, N.~Armesto, M.~Botje, M.~Cacciari, J.~Campbell and C.~Carli {\it et al.},
  %``Proton-Nucleus Collisions at the LHC: Scientific Opportunities and Requirements,''
  J.\ Phys.\ G {\bf 39} (2012) 015010
  [arXiv:1105.3919 [hep-ph]].
  %%CITATION = ARXIV:1105.3919;%%
  %51 citations counted in INSPIRE as of 10 Jan 2014aua

%\cite{Hirai:2007sx}
\bibitem{Hirai:2007sx}
  M.~Hirai, S.~Kumano and T.~-H.~Nagai,
  %``Determination of nuclear parton distribution functions and their uncertainties in next-to-leading order,''
  Phys.\ Rev.\ C {\bf 76} (2007) 065207
  [arXiv:0709.3038 [hep-ph]].
  %%CITATION = ARXIV:0709.3038;%%
  %155 citations counted in INSPIRE as of 14 Nov 2013

  %\cite{Eskola:2009uj}
\bibitem{Eskola:2009uj}
  K.~J.~Eskola, H.~Paukkunen and C.~A.~Salgado,
  %``EPS09: A New Generation of NLO and LO Nuclear Parton Distribution Functions,''
  JHEP {\bf 0904} (2009) 065
  [arXiv:0902.4154 [hep-ph]].
  %%CITATION = ARXIV:0902.4154;%%
  %292 citations counted in INSPIRE as of 14 Nov 2013

%\cite{deFlorian:2011fp}
\bibitem{deFlorian:2011fp}
  D.~de Florian, R.~Sassot, P.~Zurita and M.~Stratmann,
  %``Global Analysis of Nuclear Parton Distributions,''
  Phys.\ Rev.\ D {\bf 85} (2012) 074028
  [arXiv:1112.6324 [hep-ph]].
  %%CITATION = ARXIV:1112.6324;%%
  %38 citations counted in INSPIRE as of 14 Nov 2013

%\cite{Schienbein:2009kk}
\bibitem{Schienbein:2009kk}
  I.~Schienbein, J.~Y.~Yu, K.~Kovarik, C.~Keppel, J.~G.~Morfin, F.~Olness and J.~F.~Owens,
  %``PDF Nuclear Corrections for Charged and Neutral Current Processes,''
  Phys.\ Rev.\ D {\bf 80} (2009) 094004
  [arXiv:0907.2357 [hep-ph]].
  %%CITATION = ARXIV:0907.2357;%%
  %68 citations counted in INSPIRE as of 14 Nov 2013

%\cite{Kovarik:2010uv}
\bibitem{Kovarik:2010uv}
  K.~Kovarik, I.~Schienbein, F.~I.~Olness, J.~Y.~Yu, C.~Keppel, J.~G.~Morfin, J.~F.~Owens and T.~Stavreva,
  %``Nuclear corrections in neutrino-nucleus DIS and their compatibility with global NPDF analyses,''
  Phys.\ Rev.\ Lett.\  {\bf 106} (2011) 122301
  [arXiv:1012.0286 [hep-ph]].
  %%CITATION = ARXIV:1012.0286;%%
  %55 citations counted in INSPIRE as of 14 Nov 2013

%\cite{Kovarik:2013sya}
\bibitem{Kovarik:2013sya}
  K.~Kovarik, T.~Jezo, A.~Kusina, F.~I.~Olness, I.~Schienbein, T.~Stavreva and J.~Y.~Yu,
  %``CTEQ nuclear parton distribution functions,''
  arXiv:1307.3454 [hep-ph].
  %%CITATION = ARXIV:1307.3454;%%
  %2 citations counted in INSPIRE as of 14 Nov 2013

%\cite{Helenius:2012wd}
\bibitem{Helenius:2012wd}
  I.~Helenius, K.~J.~Eskola, H.~Honkanen and C.~A.~Salgado,
  %``Impact-Parameter Dependent Nuclear Parton Distribution Functions: EPS09s and EKS98s and Their Applications in Nuclear Hard Processes,''
  JHEP {\bf 1207} (2012) 073
  [arXiv:1205.5359 [hep-ph]].
  %%CITATION = ARXIV:1205.5359;%%
  %28 citations counted in INSPIRE as of 06 Jan 2014
  
%\cite{Martin:2004dh}
\bibitem{Martin:2004dh}
  A.~D.~Martin, R.~G.~Roberts, W.~J.~Stirling and R.~S.~Thorne,
  %``Parton distributions incorporating QED contributions,''
  Eur.\ Phys.\ J.\ C {\bf 39} (2005) 155
  [hep-ph/0411040].
  %%CITATION = HEP-PH/0411040;%%
  %237 citations counted in INSPIRE as of 14 Nov 2013

%\cite{Ball:2013hta}
\bibitem{Ball:2013hta}
  R.~D.~Ball {\it et al.}  [The NNPDF Collaboration],
  %``Parton distributions with QED corrections,''
  Nucl.\ Phys.\ B {\bf 877} (2013) 2,  290
  [arXiv:1308.0598 [hep-ph]].
  %%CITATION = ARXIV:1308.0598;%%
  %11 citations counted in INSPIRE as of 14 Nov 2013

%\cite{Bertone:2013vaa}
\bibitem{Bertone:2013vaa}
  V.~Bertone, S.~Carrazza and J.~Rojo,
  %``APFEL: A PDF Evolution Library with QED corrections,''
  arXiv:1310.1394 [hep-ph].
  %%CITATION = ARXIV:1310.1394;%%
  %1 citations counted in INSPIRE as of 14 Nov 2013

  %\cite{Sadykov:2014aua}
\bibitem{Sadykov:2014aua}
  R.~Sadykov,
  %``Impact of QED radiative corrections on Parton Distribution Functions,''
  arXiv:1401.1133 [hep-ph].
  %%CITATION = ARXIV:1401.1133;%%

  %\cite{Martin:2012da}
\bibitem{Martin:2012da}
  A.~D.~Martin, A.~J.~T.~.M.~Mathijssen, W.~J.~Stirling, R.~S.~Thorne, B.~J.~A.~Watt and G.~Watt,
  %``Extended Parameterisations for MSTW PDFs and their effect on Lepton Charge Asymmetry from W Decays,''
  Eur.\ Phys.\ J.\ C {\bf 73} (2013) 2318
  [arXiv:1211.1215 [hep-ph]].
  %%CITATION = ARXIV:1211.1215;%%
  %16 citations counted in INSPIRE as of 06 Jan 2014

%\cite{deFlorian:2003qf}
\bibitem{deFlorian:2003qf}
  D.~de Florian and R.~Sassot,
  %``Nuclear parton distributions at next-to-leading order,''
  Phys.\ Rev.\ D {\bf 69} (2004) 074028
  [hep-ph/0311227].
  %%CITATION = HEP-PH/0311227;%%
  %239 citations counted in INSPIRE as of 14 Nov 2013

  %\cite{Abelev:2009hx}
\bibitem{Abelev:2009hx}
  B.~I.~Abelev {\it et al.}  [STAR Collaboration],
  %``Inclusive $\pi^0$, $\eta$, and direct photon production at high transverse momentum in $p+p$ and $d+$Au collisions at $\sqrt{s_{NN}}=200$ GeV,''
  Phys.\ Rev.\ C {\bf 81} (2010) 064904
  [arXiv:0912.3838 [hep-ex]].
  %%CITATION = ARXIV:0912.3838;%%
  %12 citations counted in INSPIRE as of 03 Jul 2013ei
  
%\cite{Adler:2006wg}
\bibitem{Adler:2006wg}
  S.~S.~Adler {\it et al.}  [PHENIX Collaboration],
  %``Centrality dependence of pi0 and eta production at large transverse momentum in s(NN)**(1/2) = 200-GeV d+Au collisions,''
  Phys.\ Rev.\ Lett.\  {\bf 98} (2007) 172302
  [nucl-ex/0610036].
  %%CITATION = NUCL-EX/0610036;%%
  %96 citations counted in INSPIRE as of 03 Jul 2013

%\cite{Eskola:2012rg}
\bibitem{Eskola:2012rg}
  K.~J.~Eskola,
  %``Global analysis of nuclear PDFs - latest developments,''
  Nucl.\  Phys.\ A {\bf  910-911} (2013) 163
  [arXiv:1209.1546 [hep-ph]].
  %%CITATION = ARXIV:1209.1546;%%
  %4 citations counted in INSPIRE as of 06 Jan 2014

  %\cite{Kniehl:2000fe}
\bibitem{Kniehl:2000fe}
  B.~A.~Kniehl, G.~Kramer and B.~Potter,
  %``Fragmentation functions for pions, kaons, and protons at next-to-leading order,''
  Nucl.\ Phys.\ B {\bf 582} (2000) 514
  [hep-ph/0010289].
  %%CITATION = HEP-PH/0010289;%%
  %412 citations counted in INSPIRE as of 14 Nov 2013
  
%\cite{Sassot:2009sh}
\bibitem{Sassot:2009sh}
  R.~Sassot, M.~Stratmann and P.~Zurita,
  %``Fragmentations Functions in Nuclear Media,''
  Phys.\ Rev.\ D {\bf 81} (2010) 054001
  [arXiv:0912.1311 [hep-ph]].
  %%CITATION = ARXIV:0912.1311;%%
  %17 citations counted in INSPIRE as of 14 Nov 2013

  %\cite{Schienbein:2007fs}
\bibitem{Schienbein:2007fs}
  I.~Schienbein {\it et al.},
  %, J.~Y.~Yu, C.~Keppel, J.~G.~Morfin, F.~Olness and J.~F.~Owens,
  %``Nuclear parton distribution functions from neutrino deep inelastic scattering,''
  Phys.\ Rev.\ D {\bf 77} (2008) 054013
  [arXiv:0710.4897 [hep-ph]].
  %%CITATION = ARXIV:0710.4897;%%
  %61 citations counted in INSPIRE as of 14 Nov 2013

%\cite{Tzanov:2005kr}
\bibitem{Tzanov:2005kr}
  M.~Tzanov {\it et al.}  [NuTeV Collaboration],
  %``Precise measurement of neutrino and anti-neutrino differential cross sections,''
  Phys.\ Rev.\ D {\bf 74} (2006) 012008
  [hep-ex/0509010].
  %%CITATION = HEP-EX/0509010;%%
  %97 citations counted in INSPIRE as of 14 Nov 2013

%\cite{Paukkunen:2010hb}
\bibitem{Paukkunen:2010hb}
  H.~Paukkunen and C.~A.~Salgado,
  %``Compatibility of neutrino DIS data and global analyses of parton distribution functions,''
  JHEP {\bf 1007} (2010) 032
  [arXiv:1004.3140 [hep-ph]].
  %%CITATION = ARXIV:1004.3140;%%
  %25 citations counted in INSPIRE as of 14 Nov 2013

%\cite{Onengut:2005kv}
\bibitem{Onengut:2005kv}
  G.~Onengut {\it et al.}  [CHORUS Collaboration],
  %``Measurement of nucleon structure functions in neutrino scattering,''
  Phys.\ Lett.\ B {\bf 632} (2006) 65.
  %%CITATION = PHLTA,B632,65;%%
  %63 citations counted in INSPIRE as of 14 Nov 2013

%\cite{Berge:1989hr}
\bibitem{Berge:1989hr}
  J.~P.~Berge, H.~Burkhardt, F.~Dydak, R.~Hagelberg, M.~Krasny, H.~J.~Meyer, P.~Palazzi and F.~Ranjard {\it et al.},
  %``A Measurement of Differential Cross-Sections and Nucleon Structure Functions in Charged Current Neutrino Interactions on Iron,''
  Z.\ Phys.\ C {\bf 49} (1991) 187.
  %%CITATION = ZEPYA,C49,187;%%
  %222 citations counted in INSPIRE as of 14 Nov 2013

%\cite{Paukkunen:2013grz}
\bibitem{Paukkunen:2013grz}
  H.~Paukkunen and C.~A.~Salgado,
  %``Agreement of Neutrino Deep Inelastic Scattering Data with Global Fits of Parton Distributions,''
  Phys.\ Rev.\ Lett.\  {\bf 110} (2013) 212301
  [arXiv:1302.2001 [hep-ph]].
  %%CITATION = ARXIV:1302.2001;%%
  %6 citations counted in INSPIRE as of 14 Nov 2013

  %\cite{Chatrchyan:2011wt}
\bibitem{Chatrchyan:2011wt}
  S.~Chatrchyan {\it et al.}  [CMS Collaboration],
  %``Measurement of the Rapidity and Transverse Momentum Distributions of $Z$ Bosons in $pp$ Collisions at $\sqrt{s}=7$ TeV,''
  Phys.\ Rev.\ D {\bf 85} (2012) 032002
  [arXiv:1110.4973 [hep-ex]].
  %%CITATION = ARXIV:1110.4973;%%
  %51 citations counted in INSPIRE as of 06 Jan 2014

%\cite{Abazov:2007jy}
\bibitem{Abazov:2007jy}
  V.~M.~Abazov {\it et al.}  [D0 Collaboration],
  %``Measurement of the shape of the boson rapidity distribution for $p \bar{p} \to Z/gamma^* \to e^{+} e^{-}$ + $X$ events produced at $\sqrt{s}$ of 1.96-TeV,''
  Phys.\ Rev.\ D {\bf 76} (2007) 012003
  [hep-ex/0702025 [HEP-EX]].
  %%CITATION = HEP-EX/0702025;%%
  %79 citations counted in INSPIRE as of 06 Jan 2014

%\cite{Ball:2010gb}
\bibitem{Ball:2010gb}
  R.~D.~Ball {\it et al.}  [NNPDF Collaboration],
  %``Reweighting NNPDFs: the W lepton asymmetry,''
  Nucl.\ Phys.\ B {\bf 849} (2011) 112
   [Erratum-ibid.\ B {\bf 854} (2012) 926]
   [Erratum-ibid.\ B {\bf 855} (2012) 927]
  [arXiv:1012.0836 [hep-ph]].
  %%CITATION = ARXIV:1012.0836;%%
  %43 citations counted in INSPIRE as of 20 Nov 2013

%\cite{Ball:2011gg}
\bibitem{Ball:2011gg}
  R.~D.~Ball, V.~Bertone, F.~Cerutti, L.~Del Debbio, S.~Forte, A.~Guffanti, N.~P.~Hartland and J.~I.~Latorre {\it et al.},
  %``Reweighting and Unweighting of Parton Distributions and the LHC W lepton asymmetry data,''
  Nucl.\ Phys.\ B {\bf 855} (2012) 608
  [arXiv:1108.1758 [hep-ph]].
  %%CITATION = ARXIV:1108.1758;%%
  %46 citations counted in INSPIRE as of 06 Jan 2014

%\cite{Watt:2012tq}
\bibitem{Watt:2012tq}
  G.~Watt and R.~S.~Thorne,
  %``Study of Monte Carlo approach to experimental uncertainty propagation with MSTW 2008 PDFs,''
  JHEP {\bf 1208} (2012) 052
  [arXiv:1205.4024 [hep-ph]].
  %%CITATION = ARXIV:1205.4024;%%
  %20 citations counted in INSPIRE as of 20 Nov 2013
 
  %\cite{Armesto:2013kqa}
\bibitem{Armesto:2013kqa}
  N.~Armesto, J.~Rojo, C.~A.~Salgado and P.~Zurita,
  %``Bayesian reweighting of nuclear PDFs and constraints from proton-lead collisions at the LHC,''
  JHEP {\bf 1311} (2013) 015
  [arXiv:1309.5371 [hep-ph]].
  %%CITATION = ARXIV:1309.5371;%%

  %\cite{Kang:2012am}
\bibitem{Kang:2012am}
  Z.~-B.~Kang and J.~-W.~Qiu,
  %``Nuclear modification of vector boson production in proton-lead collisions at the LHC,''
  Phys.\ Lett.\ B {\bf 721} (2013) 277
  [arXiv:1212.6541].
  %%CITATION = ARXIV:1212.6541;%%
  %9 citations counted in INSPIRE as of 01 Apr 2014

%\cite{Brandt:2014vva}
\bibitem{Brandt:2014vva}
  M.~Brandt, M.~Klasen and F.~König,
  %``Nuclear parton density modifications from low-mass lepton pair production at the LHC,''
  arXiv:1401.6817 [hep-ph].
  %%CITATION = ARXIV:1401.6817;%%
  %1 citations counted in INSPIRE as of 01 Apr 2014

%\cite{Guzey:2012jp}
\bibitem{Guzey:2012jp}
  V.~Guzey, M.~Guzzi, P.~M.~Nadolsky, M.~Strikman and B.~Wang,
  %``Massive neutral gauge boson production as a probe of nuclear modifications of parton distributions at the LHC,''
  Eur.\ Phys.\ J.\ A {\bf 49} (2013) 35
   [Eur.\ Phys.\ J.\ A {\bf 49} (2013) 35]
  [arXiv:1212.5344 [hep-ph]].
  %%CITATION = ARXIV:1212.5344;%%
  %2 citations counted in INSPIRE as of 01 Apr 2014
  
%\cite{Paukkunen:2010qg}
\bibitem{Paukkunen:2010qg}
  H.~Paukkunen and C.~A.~Salgado,
  %``Constraints for the nuclear parton distributions from Z and W production at the LHC,''
  JHEP {\bf 1103} (2011) 071
  [arXiv:1010.5392 [hep-ph]].
  %%CITATION = ARXIV:1010.5392;%%
  %27 citations counted in INSPIRE as of 01 Apr 2014

%\cite{Arleo:2007js}
\bibitem{Arleo:2007js}
  F.~Arleo and T.~Gousset,
  %``Measuring gluon shadowing with prompt photons at RHIC and LHC,''
  Phys.\ Lett.\ B {\bf 660} (2008) 181
  [arXiv:0707.2944 [hep-ph]].
  %%CITATION = ARXIV:0707.2944;%%
  %23 citations counted in INSPIRE as of 01 Apr 2014
 
%\cite{Stavreva:2010mw}
\bibitem{Stavreva:2010mw}
  T.~Stavreva, I.~Schienbein, F.~Arleo, K.~Kovarik, F.~Olness, J.~Y.~Yu and J.~F.~Owens,
  %``Probing gluon and heavy-quark nuclear PDFs with gamma + Q production in pA collisions,''
  JHEP {\bf 1101} (2011) 152
  [arXiv:1012.1178 [hep-ph]].
  %%CITATION = ARXIV:1012.1178;%%
  %24 citations counted in INSPIRE as of 01 Apr 2014
 
%\cite{Arleo:2011gc}
\bibitem{Arleo:2011gc}
  F.~Arleo, K.~J.~Eskola, H.~Paukkunen and C.~A.~Salgado,
  %``Inclusive prompt photon production in nuclear collisions at RHIC and LHC,''
  JHEP {\bf 1104} (2011) 055
  [arXiv:1103.1471 [hep-ph]].
  %%CITATION = ARXIV:1103.1471;%%
  %30 citations counted in INSPIRE as of 01 Apr 2014

%\cite{Albacete:2013ei}
\bibitem{Albacete:2013ei}
  J.~L.~Albacete, N.~Armesto, R.~Baier, G.~G.~Barnafoldi, J.~Barrette, S.~De, W.~-T.~Deng and A.~Dumitru {\it et al.},
  %``Predictions for $p+$Pb Collisions at sqrt s_NN = 5 TeV,''
  Int.\ J.\ Mod.\ Phys.\ E {\bf 22} (2013) 1330007
  [arXiv:1301.3395 [hep-ph]].
  %%CITATION = ARXIV:1301.3395;%%
  %31 citations counted in INSPIRE as of 01 Apr 2014

 %\cite{He:2011sg}
\bibitem{He:2011sg}
  Y.~He, B.~-W.~Zhang and E.~Wang,
  %``Cold Nuclear Matter Effects on Dijet Productions in Relativistic Heavy-ion Reactions at LHC,''
  Eur.\ Phys.\ J.\ C {\bf 72} (2012) 1904
  [arXiv:1110.6601 [hep-ph]].
  %%CITATION = ARXIV:1110.6601;%%
  %8 citations counted in INSPIRE as of 01 Apr 2014

%\cite{Dai:2013xca}
\bibitem{Dai:2013xca}
  W.~Dai, S.~-Y.~Chen, B.~-W.~Zhang and E.~-K.~Wang,
  %``Cold nuclear matter effects on isolated prompt photon and isolated prompt photon+jet productions in relativistic heavy-ion collisions,''
  Commun.\ Theor.\ Phys.\  {\bf 59} (2013) 349.
  %%CITATION = CTPMD,59,349;%%


  %\cite{Chatrchyan:2014hqa}
\bibitem{Chatrchyan:2014hqa}
  S.~Chatrchyan {\it et al.}  [CMS Collaboration],
  %``Studies of dijet pseudorapidity distributions and transverse momentum balance in pPb collisions at $\sqrt{s_{NN}}$=5.02 TeV,''
  arXiv:1401.4433 [nucl-ex].
  %%CITATION = ARXIV:1401.4433;%%
  %1 citations counted in INSPIRE as of 02 Apr 2014
  
%\cite{Eskola:2013aya}
\bibitem{Eskola:2013aya}
  K.~J.~Eskola, H.~Paukkunen and C.~A.~Salgado,
  %``A perturbative QCD study of dijets in p+Pb collisions at the LHC,''
  JHEP {\bf 1310} (2013) 213
  [arXiv:1308.6733 [hep-ph]].
  %%CITATION = ARXIV:1308.6733;%%
  %3 citations counted in INSPIRE as of 02 Apr 2014

%\cite{Chatrchyan:2012vq}
\bibitem{Chatrchyan:2012vq}
  S.~Chatrchyan {\it et al.}  [CMS Collaboration],
  %``Measurement of isolated photon production in $pp$ and PbPb collisions at $\sqrt{s_{NN}}=2.76$ TeV,''
  Phys.\ Lett.\ B {\bf 710} (2012) 256
  [arXiv:1201.3093 [nucl-ex]].
  %%CITATION = ARXIV:1201.3093;%%
  %45 citations counted in INSPIRE as of 01 Apr 2014

%\cite{Chatrchyan:2012nt}
\bibitem{Chatrchyan:2012nt}
  S.~Chatrchyan {\it et al.}  [CMS Collaboration],
  %``Study of $W$ boson production in PbPb and $pp$ collisions at $\sqrt{s_{NN}}=2.76$ TeV,''
  Phys.\ Lett.\ B {\bf 715} (2012) 66
  [arXiv:1205.6334 [nucl-ex]].
  %%CITATION = ARXIV:1205.6334;%%
  %26 citations counted in INSPIRE as of 01 Apr 2014

%\cite{Chatrchyan:2011ua}
\bibitem{Chatrchyan:2011ua}
  S.~Chatrchyan {\it et al.}  [CMS Collaboration],
  %``Study of Z boson production in PbPb collisions at nucleon-nucleon centre of mass energy = 2.76 TeV,''
  Phys.\ Rev.\ Lett.\  {\bf 106} (2011) 212301
  [arXiv:1102.5435 [nucl-ex]].
  %%CITATION = ARXIV:1102.5435;%%
  %54 citations counted in INSPIRE as of 01 Apr 2014

%\cite{Aad:2012ew}
\bibitem{Aad:2012ew}
  G.~Aad {\it et al.}  [ATLAS Collaboration],
  %``Measurement of $Z$ boson Production in Pb+Pb Collisions at $\sqrt{s_{NN}}=2.76$ TeV with the ATLAS Detector,''
  Phys.\ Rev.\ Lett.\  {\bf 110} (2013) 022301
  [arXiv:1210.6486 [hep-ex]].
  %%CITATION = ARXIV:1210.6486;%%
  %13 citations counted in INSPIRE as of 01 Apr 2014aua
  
%\cite{Paatelainen:2013eea}
\bibitem{Paatelainen:2013eea}
  R.~Paatelainen, K.~J.~Eskola, H.~Niemi and K.~Tuominen,
  %``Fluid dynamics with saturated minijet initial conditions in ultrarelativistic heavy-ion collisions,''
  arXiv:1310.3105 [hep-ph].
  %%CITATION = ARXIV:1310.3105;%%
  %4 citations counted in INSPIRE as of 01 Apr 2014

%\cite{Klasen:2013mga}
\bibitem{Klasen:2013mga}
  M.~Klasen, C.~Klein-Bösing, F.~König and J.~P.~Wessels,
  %``How robust is a thermal photon interpretation of the ALICE low-$p_{T}$ data?,''
  JHEP {\bf 1310} (2013) 119
  [arXiv:1307.7034].
  %%CITATION = ARXIV:1307.7034;%%
  %5 citations counted in INSPIRE as of 02 Apr 2014
  
%\cite{Chatterjee:2013naa}
\bibitem{Chatterjee:2013naa}
  R.~Chatterjee, H.~Holopainen, I.~Helenius, T.~Renk and K.~J.~Eskola,
  %``Elliptic flow of thermal photons from event-by-event hydrodynamic model,''
  Phys.\ Rev.\ C {\bf 88} (2013) 034901
  [arXiv:1305.6443 [hep-ph]].
  %%CITATION = ARXIV:1305.6443;%%
  %13 citations counted in INSPIRE as of 02 Apr 2014

  
\end{thebibliography}
\end{document}